# DISCUSSION OF "2004 IMS MEDALLION LECTURE: LOCAL RADEMACHER COMPLEXITIES AND ORACLE INEQUALITIES IN RISK MINIMIZATION" BY V. KOLTCHINSKII

By Peter L. Bartlett and Shahar Mendelson

*University of California at Berkeley and Australian National University*

**1. Relating empirical and real structures: additive and multiplicative results.** The key issue investigated in Vladimir Koltchinskii's paper is the behavior of an empirical minimizer $\hat{f} \in F$, that is, a function $f$ in $F$ with minimal sample average,

$$P_n f = \frac{1}{n} \sum_{i=1}^{n} f(X_i),$$

where $X_1, \ldots, X_n$ are drawn i.i.d. from a probability measure $P$ on $\mathcal{X}$ and $F$ is a class of real-valued functions defined on $\mathcal{X}$. The study of bounds on the expectation $P\hat{f}$ arises in many applied areas, including the analysis of randomized optimization methods involving Monte Carlo estimates of integrals. Motivated by prediction problems that arise in machine learning and nonparametric statistics, the paper makes an important contribution to the study of these bounds, and to the development of model selection methods that exploit the bounds.

The broad approach taken in this paper, and in much earlier work, is to show that the empirical structure (i.e., the collection of sample averages, $P_n f$) is close to the real structure (i.e., the collection of expectations, $Pf$). If they are close in the additive sense that $\|P_n - P\|_F$ decreases at some rate, then it is clear that $P\hat{f}$ approaches $\inf_{f \in F} Pf$ at that rate. As the paper recalls, there is a tight relationship between the Rademacher process indexed by coordinate projections of the class $F$ and this additive notion of closeness of empirical and real structures. Also, it can be advantageous to consider these properties only locally, that is, in the set $F(\delta) \subset F$ of near-minimizers of $Pf$. In particular, if the variance of elements of $F(\delta)$ goes to zero with $\delta$, then faster rates are possible through the study of these local properties.

---

Received March 2006.

An alternative, developed in the paper, is closeness in the multiplicative sense that for $0 < \varepsilon < 1$, for all functions $f$ in $F$ that have expectations not too small,

$$(1 - \varepsilon)P_n f \leq Pf \leq (1 + \varepsilon)P_n f.$$

Again, these results rely on the variance of an element of $F$ decreasing as its expectation decreases. Let us call a class that has this property a Bernstein class.

DEFINITION 1.1. We say that $F$ is a $(\beta, B)$-Bernstein class with respect to the probability measure $P$ (where $0 < \beta \leq 1$ and $B \geq 1$), if every $f$ in $F$ satisfies

$$Pf^2 \leq B(Pf)^\beta.$$

This condition arises naturally in many situations, as the paper describes. Obviously, if $F$ consists of nonnegative functions bounded by $b$, then $F$ is a Bernstein class (with $\beta = 1$) with respect to any probability measure. Other examples arise for excess loss classes,

$$F = \{\ell_g - \ell_{g^*} : g \in G\} \qquad \text{with } \ell_g(x, y) = \ell(g(x), y),$$

where $\ell : \mathbb{R}^2 \to [0, \infty)$ is a loss function and $g^* \in G$ minimizes $P\ell_g$. For example, in regression, if $\alpha \mapsto \ell(\alpha, y)$ are uniformly convex Lipschitz bounded functions and $G$ is convex, then $F$ is a Bernstein class [2, 5, 6]. In pattern classification with $\ell$ the discrete loss, if $g^*$ is the Bayes rule and the conditional probability $\Pr(Y = 1 | X)$ is unlikely to be near $1/2$, then the excess loss class is Bernstein [10].

Under some mild assumptions on a Bernstein class $F$, there is a simple proof of the multiplicative closeness of the empirical and true structures, using Talagrand-style concentration inequalities for empirical processes [9]. The assumptions are that functions in $F$ are bounded, and that $F$ is star-shaped around 0, that is, for every $0 \leq a \leq 1$ and any $f \in F$, $af \in F$. A generalization of the following result (for arbitrary Bernstein conditions) appears in [3].

THEOREM 1.2. *There exists an absolute constant $c$ for which the following holds. For $F$ a $(1, B)$-Bernstein class of functions bounded by $b$ which is star-shaped around 0, with probability at least $1 - e^{-x}$, the empirical minimizer $\hat{f} \in F$ satisfies*

$$P\hat{f} \leq \max\left\{\inf\{r > 0 : \xi_n(r) \leq r/4\}, \frac{c(b + B)x}{n}\right\},$$

*where*

$$\xi_n(r) = \mathbb{E} \sup\{Pf - P_n f : f \in F_r\} \qquad \text{with } F_r = \{f \in F : Pf = r\}.$$

The proof uses a simple geometric argument: Talagrand's inequality implies that, for the subset $F_r$ with $r$ not too small, there is a near-equivalence between the multiplicative comparison inequality

$$(1 - \varepsilon)P_n f \leq Pf \leq (1 + \varepsilon)P_n f$$

holding uniformly over $F_r$, and the expectation of the supremum of the empirical process, $\mathbb{E}\|P - P_n\|_{F_r}$ being less than $r\varepsilon$. And then the star-shaped property shows that this extends to all functions in $F$ that have $Pf \geq r$. The reason that the critical level $r$ cannot be too small is because by the star-shape property, the "relative complexity" of the sets $F_r$ increases as $r$ decreases.

Notice that this result is in terms of the fixed point

$$\inf\{r > 0 : \xi_n(r) \leq r/4\}$$

which is never larger than fixed points of the related functions in Koltchinskii's paper. In particular, $\xi_n(r)$ is bounded by $\mathbb{E}\|P - P_n\|_{F_r}$, the expected supremum of the empirical process indexed by functions that have expectation $r$, whereas the paper considers expected suprema over larger sets, defined by the $L_2(P)$ structure.

## 2. Data-dependent bounds and model selection.

One of the appealing features of these kinds of bounds, which is developed in Koltchinskii's paper, is that there are empirical versions that show that we can accurately estimate the bounds using the sample. It turns out that this is also the case for the result of Theorem 1.2; see [4]. The idea is to replace the quantity $\xi_n(r) = \mathbb{E}\|P - P_n\|_{F_r}$ with a sample-based estimate of the corresponding Rademacher averages,

$$\hat{\xi}_n(r) = R_n(\hat{F}_r) \qquad \text{with } \hat{F}_r = \{f \in F : c_1 r \leq P_n f \leq c_2 r\},$$

for some constants $c_1 < 1 < c_2$. The same concentration properties that imply the bounds in terms of the fixed point of $\xi_n(r)$ show that a fixed point of $\hat{\xi}_n(r) + c_3 r$ also suffices.

Another interesting contribution in the paper is the application of these bounds in terms of empirical quantities to model selection problems. It is natural to consider how estimates of expectations (i.e., estimates of risk, in the case of loss classes) can be used to define penalization methods for model selection. In particular, define the risk $P\ell_f = P\ell(Y, f(X))$, where $\ell$ is a nonnegative *loss function* and $(X, Y)$ is a covariate/response pair. Suppose that we have a sequence $F_1, F_2, \ldots$ of function classes defined on $\mathcal{X}$, and we use an estimator that first chooses the empirical minimizer

$$\hat{f}_k = \operatorname*{arg\,min}_{f \in F_k} P_n \ell_f,$$

from each $F_k$, and then picks $\hat{f} = \hat{f}_{\hat{k}}$ as the $\hat{f}_k$ that minimizes a penalized risk of the form

$$P_n \ell_{\hat{f}_k} + \hat{p}(k).$$

A key concern in these problems is proving oracle inequalities of the form

$$\Pr\left(P\ell_{\hat{f}} \geq \inf_{k \geq 1}\left\{\inf_{f \in F_k} P\ell_f + p(k)\right\}\right) \to 0,$$

where $p(k)$ is a complexity penalty related to the penalty $\hat{p}(k)$ used by the method. Notice, in particular, the constant multiplying the risk $(P\ell_f)$ term is 1. It turns out that if the classes are ordered by inclusion, then multiplicative bounds for the excess loss class immediately give such oracle inequalities. The multiplicative bounds we need are of the form

$$\forall f \in F, \qquad (1 - \varepsilon)P_n f - r \leq Pf \leq (1 + \varepsilon)P_n f + r.$$

(Notice that, although upper bounds of this kind are immediate from the proof of Theorem 1.2, the lower bounds are not.) The following theorem is elementary; it is proved in [1]. Define $f_k^*$ as the element of $F_k$ that minimizes $P\ell_f$.

THEOREM 2.1. *Suppose that*

$$\sup_k \sup_{f \in F_k} (P\ell_f - P\ell_{f_k^*} - 2(P_n \ell_f - P_n \ell_{f_k^*}) - \varepsilon_k) \leq 0,$$

$$\sup_k \sup_{f \in F_k} (P_n \ell_f - P_n \ell_{f_k^*} - 2(P\ell_f - P\ell_{f_k^*}) - \varepsilon_k) \leq 0,$$

*where the classes are ordered by inclusion, and the quantities $\varepsilon_k$ are similarly ordered, $F_1 \subseteq F_2 \subseteq F_3 \subseteq \cdots$, $\varepsilon_1 \leq \varepsilon_2 \leq \varepsilon_3 \leq \cdots$. Then choosing $p(k) = 7\varepsilon_k/2$ ensures that*

$$P\ell_{\hat{f}} \leq \inf_k (P\ell_{f_k^*} + 9\varepsilon_k).$$

**3. Lower bounds.** It is interesting to consider the tightness of the upper bounds of the type proved in the paper. Koltchinskii provides examples that demonstrate optimal rates in several minimax settings. But is it true that, for all function classes and probability distributions, the upper bounds imply the correct rate of convergence of $P\hat{f}$ to its asymptotic value?

It turns out that they are not tight. Indeed, in attempting to prove matching lower bounds, we were led to the following theorem (see [3]), which uses a direct analysis of the empirical minimizer to give essentially matching upper and lower bounds on its expectation, in terms of a related property of the empirical process. Set

$$\xi_n(r) = \mathbb{E} \sup_{f \in F_r} Pf - P_n f \qquad \text{where } F_r = \{f \in F : Pf = r\},$$

and, for $\varepsilon > 0$, define

$$r_{\varepsilon,+} = \sup\Big\{0 \le r \le b : \xi_n(r) - r \ge \sup_s(\xi_n(s) - s) - \varepsilon\Big\},$$

$$r_{\varepsilon,-} = \inf\Big\{0 \le r \le b : \xi_n(r) - r \ge \sup_s(\xi_n(s) - s) - \varepsilon\Big\}.$$

These two quantities bracket the range of values of $r$ that $\varepsilon$-approximately maximize the function $\xi_n(r) - r$. The theorem shows that, for $\varepsilon$ not too small, they also bracket the expectation of the empirical minimizer.

THEOREM 3.1.   *For any $c_1 > 0$, there is a constant $c$ such that the following holds. Let $F$ be a $(1, B)$-Bernstein class that is star-shaped at 0. Define $s$, $r_{\varepsilon,+}$ and $r_{\varepsilon,-}$ as above, and set*

$$r' = \max\Big\{\inf\{r > 0 : \xi_n(r) \le r/4\}, \frac{c(b+B)(x+\log n)}{n}\Big\}.$$

*Let $\hat{f}$ denote an empirical risk minimizer. If*

$$\varepsilon \ge c\Big(\max\Big\{\sup_{s>0}(\xi_n(s) - s), r'\Big\}\frac{(B+b)(x+\log n)}{n}\Big)^{1/2},$$

*then*

1. *With probability at least $1 - e^{-x}$,*

$$\mathbb{E}\hat{f} \le \max\Big\{\frac{1}{n}, r_{\varepsilon,+}\Big\}.$$

2. *If*

$$\xi_n(0, c_1/n) < \sup_{s>0}(\xi_n(s) - s) - \varepsilon,$$

*then with probability at least $1 - e^{-x}$,*

$$\mathbb{E}\hat{f} \ge r_{\varepsilon,-}.$$

The following theorem (see [3]) shows that there is a real gap between this result and the bounds in terms of fixed points of $\xi_n(r)$ described in Theorem 1.2 and thus between this result and the similar bounds in Koltchinskii's paper.

THEOREM 3.2.   *There is an absolute constant $c$ for which the following holds. If $0 < \delta < 1$ and $n > N_0(\delta)$, there is a probability measure $P$ and a star-shaped class $F$, which consists of functions bounded by 1 and is a $(1, 2)$-Bernstein class, such that*

1. *For every $X_1, \ldots, X_n$ there is a function $f \in F$ with $\mathbb{E}f = 1/4$ and $\mathbb{E}_n f = 0$.*

2. *For the class $F$, $\inf\{r > 0 : \xi_n(r) \leq r/4\} = 1/4$.*

3. *If $\hat{f}$ is a $\rho$-approximate empirical minimizer, where $0 < \rho < 1/8$, then with probability larger than $1 - \delta$,*

$$\frac{1}{n}\left(1 - c\sqrt{\frac{\log n}{n}} - \rho\right) \leq \mathbb{E}\hat{f} \leq \frac{1}{n}.$$

So there is an example in which Theorem 3.1 demonstrates that $P\hat{f}$ is of order $1/n$, but the local Rademacher bounds are constants. Although the example is of a class $F$, it is straightforward to show that, under mild conditions on a loss function $\ell$, this class can be written as an excess loss class $\{\ell_g - \ell_{g^*} : g \in G\}$ for some $G$ and some probability distribution (see [3]).

We have seen that we can obtain a data-dependent version of the local Rademacher bounds that can be used as complexity penalties in model selection methods. If the same thing were true for the bounds of Theorem 3.1, we could improve on these model selection methods. Unfortunately, this is not possible if one only has access to function values on finite samples. There is an example in [4] that shows that it is impossible to establish a data-dependent upper bound on the expectation of the empirical minimizer that is asymptotically better than the fixed point of $\xi_n(r)$. The idea is to construct two classes of functions that look identical when projected on any sample of finite size, but for one class both a typical expectation of the empirical minimizer and the fixed point of $\xi_n(r)$ are of the order of a constant, while for the other a typical expectation is of the order of $1/n$.

**4. The role of concentration.** Arguably, the most important contribution to modern prediction bound techniques is Talagrand's concentration inequality for empirical processes [9]. However, it is important to note that its full strength is rarely used.

Roughly speaking, this inequality ensures that with high probability, the dominant terms in the upper and lower estimates on $\|P_n - P\|_F$ are $(1 + \alpha)\mathbb{E}\|P_n - P\|_F$ and $(1 - \alpha)\mathbb{E}\|P_n - P\|_F$, where $\alpha$ can be made arbitrarily close to 0, at a price of larger second-order terms. In fact, in the vast majority of results one can take $\alpha$ to be any fixed constant $0 < \alpha < 1$.

The important point is that in multiplicative-type results (e.g., ratio-limit theorems as presented in the paper or similar to Theorem 1.2), the role of this coefficient is not important. It is only when one wishes to analyze the behavior of the empirical minimizer on the set $F_r$ and compare it to its behavior on $F_s$ for $r \neq s$ that the exact dependency on $\alpha$ is required. This is the case in the proof of Theorem 3.1.

Moreover, in the vast majority of results that do not involve multiclass analysis, the actual role of Talagrand's concentration inequality is restricted to ensuring a better dependency on the confidence level $\delta$ — from polynomial in $1/\delta$ to logarithmic in $1/\delta$. Indeed, an almost identical result to Theorem 1.2 can be proved without Talagrand's inequality, leading to the same order of error rates but with a worst constant. The dominant term remains the same—the fixed point of the function $\mathbb{E}\|P_n - P\|_{F_r}$.

One should ask: why not always use Talagrand's inequality? The reason is that it is not always available. Concentration of the supremum of an empirical process is known for a class with a bounded diameter in $L_\infty$. Thus, any result which is truly based on this concentration does not extend to unbounded classes. Of course, it could be very interesting to develop a similar theory for the unbounded case.

## 5. Some questions.

1. Talagrand's concentration inequality is a "function class" version of Bernstein's inequality, with the secondary terms determined by the $L_2$ and $L_\infty$ diameters of $F$. It could be useful (and not only from the statistical point of view) to prove a concentration result with the $L_\infty$ diameter replaced by the $\psi_1$ diameter (recall that for $\alpha \geq 1$, $\|X\|_{\psi_\alpha} = \inf\{c > 0 : \mathbb{E}\exp(|X|/c) \leq 2\}$; the $\psi_1$ norm measures the subexponential decay of $X$).

2. The results in the paper are based on the behavior of the Rademacher process indexed by a random coordinate projection of $F$ (i.e., the restriction of $F$ onto a random sample). Thus, error bounds are determined using random (empirical) $\ell_2^n$ metric on coordinate projections. It should be interesting to develop a theory of learning which uses "global" metric structures. Clearly, the $L_2(P)$ one, which is the natural candidate, is too weak, for otherwise the supremum of the empirical process indexed by $F$ could be controlled in terms of the limiting Gaussian, which is not true. It is more likely that stronger metrics (e.g., the $\psi_\alpha$ metrics) will play a central role in such a development, as in [7, 8].

COMPUTER SCIENCE DIVISION
  AND DEPARTMENT OF STATISTICS
UNIVERSITY OF CALIFORNIA AT BERKELEY
367 EVANS HALL
BERKELEY, CALIFORNIA 94720
USA
E-MAIL: bartlett@cs.berkeley.edu

CENTRE FOR MATHEMATICS
  AND ITS APPLICATIONS
AUSTRALIAN NATIONAL UNIVERSITY
CANBERRA, ACT 0200
AUSTRALIA
E-MAIL: shahar.mendelson@anu.edu.au